\documentclass[prb,aps,twocolumn,longbibliography,superscriptaddress,preprintnumbers]{revtex4-2}
\usepackage[colorlinks,bookmarks=true,citecolor=blue,linkcolor=blue,urlcolor=blue]{hyperref}
\usepackage{dcolumn,graphicx,amsfonts,amsthm,bm,color,appendix,float}

\usepackage{braket}
\usepackage[normalem]{ulem}
\usepackage[version=4]{mhchem}
\usepackage{siunitx}

\usepackage{amsmath}
\usepackage{amssymb}
\usepackage{CJKutf8}
\usepackage{mathtools}
\usepackage{enumitem}
\usepackage{booktabs}
\usepackage{bbold}
\usepackage[caption=false]{subfig}

\usepackage[dvipsnames]{xcolor}
\definecolor{pal0}{rgb}{0.8941, 0.102 , 0.1098}
\definecolor{pal1}{rgb}{0.2157, 0.4941, 0.7216}
\definecolor{pal2}{rgb}{0.302 , 0.6863, 0.2902}
\definecolor{pal3}{rgb}{0.5961, 0.3059, 0.6392}
\definecolor{pal4}{rgb}{1.    , 0.498 , 0.    }

\DeclareMathOperator{\Tr}{Tr}

\newcommand{\n}[1]{\left| #1 \right|}%%adjustable-height norm shortcut
\newcommand{\setc}[2]{\left\{#1\; :\; #2 \right\}}%%set notation shortcut with a colon
\renewcommand{\v}[1]{\boldsymbol{#1}}%%shortcut to make a vector (overwrites the default command)

\newcommand{\bz}{\mathrm{BZ}}

\newcommand{\Auc}{A_{\mathrm{uc}}}
\newcommand{\ev}{\check{e}}

\newcommand{\PRLsec}[1]{\emph{#1---}}

\begin{document}

\title{
$\lambda$-Jellium Model for the Anomalous Hall Crystal
}

\author{Tomohiro Soejima (\begin{CJK*}{UTF8}{bsmi}副島智大\end{CJK*})}
\thanks{These authors contributed equally.}
\affiliation{Department of Physics, Harvard University, Cambridge, MA 02138, USA}

\author{Junkai Dong (\begin{CJK*}{UTF8}{bsmi}董焌\end{CJK*}\begin{CJK*}{UTF8}{gbsn}锴\end{CJK*})}
\thanks{These authors contributed equally.}
\affiliation{Department of Physics, Harvard University, Cambridge, MA 02138, USA}
\affiliation{Kavli Institute for Theoretical Physics, University of California, Santa Barbara, California 93106, USA}

\author{Ashvin Vishwanath}
\affiliation{Department of Physics, Harvard University, Cambridge, MA 02138, USA}

\author{Daniel E. Parker}
\affiliation{Department of Physics, University of California at San Diego, La Jolla, California 92093, USA}

\begin{abstract}
The jellium model is a paradigmatic problem in condensed matter physics, exhibiting a phase transition between metallic and Wigner crystal phases. However, its vanishing Berry curvature makes it ill-suited for studying recent experimental platforms that combine strong interactions with nontrivial quantum geometry. These experiments inspired the anomalous Hall crystal (AHC) --- a topological variant of the Wigner crystal. The AHC spontaneously breaks continuous translation symmetry but has a nonzero Chern number. In this work, we introduce  $\lambda-$jellium, a minimal extension of the two-dimensional jellium model. Its Berry curvature distribution is controlled by a single parameter, $\lambda$, where $\lambda=0$ corresponds to the standard jellium model. This setup facilitates the systematic exploration of Berry curvature's impact on electron crystallization. The phase diagram of this model, established using self-consistent Hartree Fock calculations, reveals several interesting features: (i) The AHC phase occupies a  large region of the phase diagram. (ii) Two distinct Wigner crystal phases, the latter enabled by quantum geometry, and two distinct Fermi liquid phases are present. (iii) A continuous phase transition separates the AHC and one of the WC phases. (iv) In some parts of the AHC phase, the lattice geometry is non-triangular, unlike in the classical Wigner crystal. In addition to elucidating the physics of correlated electrons with nonzero Berry curvature, we expect that the simplicity of the model makes it an excellent starting point for more advanced numerical methods.
\end{abstract}

\maketitle

In the classic problem considered by Wigner~\cite{wigner_interaction_1934}, electron crystallization driven by strong Coulomb interactions was examined within  the jellium model, in which an electron fluid  embedded in a uniform neutralizing background spontaneously breaks translation symmetry upon decreasing the density.
The two-dimensional incarnation of the jellium model, the two-dimensional electron gas (2DEG), has since been studied extensively using advanced numerical 
techniques~\cite{Giuliani_Vignale_2005,Ceperley_VMC_Ground_state,tanatar1989ground,Gori-Giorgi_Pair-distribution,De_Palo_Effects_Thickness,Varsano_Spin-polarization,Attaccalite_Correlation_Energy,Rapisarda_DMC,Drummond_Phase_Diagram,Drummond_QMC,azadi2024quantummontecarlostudy, HF_Wigner_crystal}, which revealed the importance of beyond-mean-field effects for the energy competition of the Wigner crystal (WC) phase with the Fermi liquid.

Recent experiments~\cite{lu2024fractional, xie2025tunablefractionalcherninsulators, choi2024electricfieldcontrolsuperconductivity, waters2025chern, Lu_2025} in rhombohedral multilayer graphene (RMG), where new phases including integer and fractional quantum Hall phases emerge at low electronic densities and zero magnetic field, have inspired an explosion of theoretical interest in topological crystalline phases of matter.
We, along with our collaborators and another simultaneous work, proposed~\cite{AHC1, AHC_Yahui} the possibility of an interaction-driven Chern insulator that spontaneously breaks continuous translation symmetry, which we dubbed the anomalous Hall crystal (AHC).

The complexity of the microscopic RMG Hamiltonian, consisting of ten orbitals and a morass of single particle hopping parameters, makes it difficult to distill simple physics, preventing a unified understanding despite intensive efforts~\cite{AHC_Senthil,guo2024fractional, kwan2023moir, AHC2, Zhihuan_Stability, yu2024moir,zhou2024newclassesquantumanomalous, crepel2025efficient, bernevig2025berrytrashcanmodelinteracting}. 
This has led to a flurry of interest in simplified models for AHCs~\cite{Zeng_sublattice_structure, Tan_parent_berry, Tan_FAHC}.
However, these models are either phenomenological mean-field models, or contain a large number of spinor components, thus precluding the possibility of beyond mean-field numerics. Indeed, determining the fate of the AHC in the presence of beyond-mean-field quantum fluctuations is an urgent challenge. 

To understand the universal physics of topological band minima, eventually including quantum fluctuations, one should consider the simplest possible model. To this end, we propose $\lambda$-jellium.
This two-band model endows the quadratic dispersion of the jellium model with a nontrivial skyrmionic texture in momentum space, which is responsible for its non-zero Berry curvature. The model contains two tuning parameters: $r_s$ and $\lambda$. The familiar $r_s$ is the ratio between the quadratic dispersion and Coulomb interaction terms, whereas $1/\lambda$ controls the extent of the skyrmionic texture. In the $\lambda\to 0$ limit, the model reduces to the standard two-dimensional jellium model, thus allowing us to systematically understand the effect of Berry curvature. 
Our model, given its simplicity, limited spinor dimensions, and connection to the jellium model, can be studied with beyond mean-field techniques employed such as variational and diffusion Monte-Carlo methods developed over many decades~\cite{Ceperley_VMC_Ground_state,tanatar1989ground,Gori-Giorgi_Pair-distribution,De_Palo_Effects_Thickness,Varsano_Spin-polarization,Attaccalite_Correlation_Energy,Rapisarda_DMC,Drummond_Phase_Diagram,Drummond_QMC,azadi2024quantummontecarlostudy}. 

To pave the way for these advanced methods, we perform extensive self-consistent Hartree-Fock calculations of the phase diagram of the $\lambda$-jellium model. We find that the AHC phase occupies a significant region of the phase diagram. In addition, we also find two distinct Wigner crystal phases: a conventional Wigner crystal and a ``halo Wigner crystal" with distinct symmetry properties. Two Fermi liquids, whose Fermi surfaces are circular and annular, also appear in the phase diagram. We find that the halo Wigner crystal and the annular Fermi liquid arises due to the interplay between interactions and non-trivial quantum geometry. We further find a quantum critical point between the AHC phase and the halo Wigner crystal, which is described by an emergent Dirac cone.

Finally, we find that in some parts of the AHC phase, spontaneous crystallization picks a non-triangular lattice geometry. 
This subtlety in the choice of lattice geometry already exists in the case of Wigner crystals, where the large $r_s$ classical electrostatic limit selects the triangular lattice. However, deep in the quantum regime, the preferred lattice geometry is not \textit{a priori} clear. For instance, with spinful electrons at small interaction strengths, a square antiferromagnet wins over the triangular lattice within HF~\cite{HF_Wigner_crystal}. Similarly, in $\lambda$-jellium we find that non-triangular AHC states can be energetically preferred over triangular AHC states.
This echoes earlier works on RMG, which found unit cells with multiple electrons~\cite{zhou2024newclassesquantumanomalous,waters2025chern} or non-triangular unit cells~\cite{YBK_elastic} can be stabilized.

\begin{figure}[t]
	%\centering
	\includegraphics[width=\linewidth]{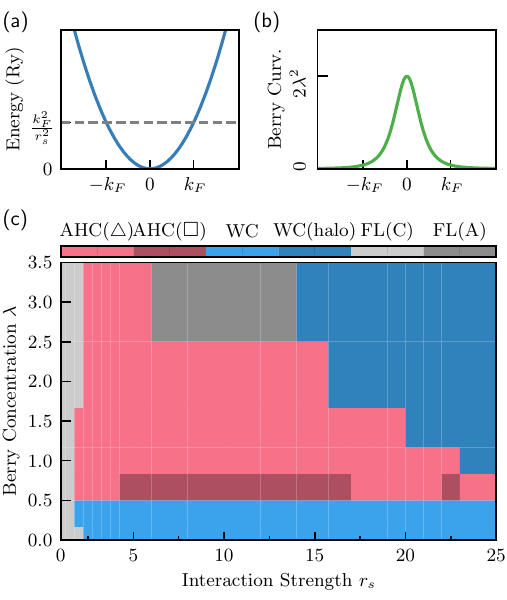}
	\caption{
    (a) Quadratic dispersion and (b) Berry curvature of the lower single-particle band of the topological electron gas model, Eq.~\eqref{eq:ham}.
	(c)~Mean-field phase diagram. The limit $\lambda = 0$ is identical to the spinless 2DEG. Two significant Fermi liquid (FL) regions are present: one at low interaction strengths, and one at large interaction strength and Berry curvature concentration. Under strong interactions, a Wigner crystal (WC) appears that undergoes a first-order transition to an anomalous Hall crystal (AHC) at $\lambda \approx \frac{1}{2}$. A putative second order transition back to a WC appears at larger $\lambda$. A significant region of AHC is present, with the competition between triangular and square AHCs shown by light pink (AHC($\triangle$)) when the triangular AHC has lower energy than the square, and dark pink (AHC($\square$)) when the reverse is true.  Within the AHC phase, a square unit cell is preferred to a triangular unit cell in a region near $\lambda = 2/3$.
    }
	\label{fig:2TEG_dispersion}
\end{figure}
\PRLsec{The $\lambda$-jellium model}
We now introduce the Hamiltonian for $\lambda$-jellium, a two-band generalization of jellium with a skyrmionic texture in momentum space that provides a tunable amount of Berry curvature.

Recall the Hamiltonian for the (spinless) 2DEG is~\cite{tanatar1989ground}
\begin{equation}
    \hat{H}_{\textsf{2DEG}} = -\frac{1}{r_s^2} \sum_{i=1}^N \nabla_i^2 + \frac{2}{r_s} \sum_{i< j}^N \frac{1}{\n{\v{r}_i - \v{r}_j}},
	\label{eq:2DEG}
\end{equation}
where length is measured in units of the typical interparticle distance $a$ (defined below), and energy has units of Rydbergs $\textrm{Ry}$~\footnote{Note that $1 \textrm{ Ry}$ differs from the other standard measure of energy, a Hartree, by a factor of two.}. These are related to microscopic units by $r_s = \frac{a}{a_0},\quad
    a = \frac{1}{\sqrt{\pi \rho}}, \quad
    a_0 = \frac{\hbar^2}{m e^2}, \quad
    \textrm{Ry} = \frac{m e^4}{2\hbar^2}$~\cite{tanatar1989ground}.
Here $\rho$ is the number density of electrons, a circle of radius $a$ encloses one particle on average, and $r_s$ is the dimensionless potential/kinetic ratio. This choice sets the Fermi momentum to be $\frac{2}{a}$, which means $k_F = 2$ in the unit system used in Eq.~\eqref{eq:2DEG}.

We now present a \textit{local} two-band generalization of this model whose lower band has identical dispersion and interaction to Eq.~\eqref{eq:2DEG}, but includes a tunable amount of Berry curvature concentrated at the band minimum.
To maintain locality, we use a kinetic energy that is a simple polynomial in the derivative operators: 
\begin{equation}
    \hat{h} = \Delta \begin{bmatrix}
        -\lambda^2\nabla^2 & i\lambda {\partial}\\
         i\lambda \bar{\partial} & 1
    \end{bmatrix},
	\label{eq:holomorphic_flat_band}
\end{equation}
where $\partial = \partial_x - i \partial_y, \nabla^2 = \partial_x^2 + \partial_y^2$ and $\Delta$ is a large and positive constant. 

One can think of this as a Dirac equation with a momentum-dependent mass (see SI~\cite{supp}). The lower band of $\hat{h}$ is exactly flat: $\epsilon_{\v{q}} = 0$. Its wavefunction is $\phi(\v{r}) = \frac{1}{\sqrt{A}} \int d^2\v{q} \; \phi_{\v{q}} \; e^{i \v{q} \cdot \v{r}}$ with spinor 
\begin{equation}
\label{eq:lowerband}
	\quad \phi_{\v{q}} = \frac{1}{\sqrt{1+\lambda^2 q^2}}\begin{bmatrix} 1\\ \lambda (q_x+i q_y) \end{bmatrix},
\end{equation}
where $q = \n{\v{q}}$.
The upper band is separated by a gap with minimum size $\Delta$. Henceforth we take $\Delta \to \infty$ and focus on the lower band.

We now study interacting electrons in the non-trivial lower band Eq.~\eqref{eq:lowerband}. To do so, 
we combine Eqs.~(\ref{eq:2DEG},\ref{eq:holomorphic_flat_band}) to form the full model, which we call $\lambda$-jellium:
\begin{equation}
\label{eq:ham}
	\hat{H}%_{\mathsf{2TEG}} %= \hat{h}_0 + \hat{H}_{\text{2DEG}}
    = \Delta \sum_{i=1}^N \begin{bmatrix}
        -\lambda^2 \nabla_i^2 & i\lambda \partial_i\\
        i\lambda \bar{\partial}_i & 1
    \end{bmatrix}
    - 
	\sum_{i=1}^N \hat{I}_2 \frac{\nabla_i^2}{r_s^2} + \frac{2}{r_s} \sum_{i<j}^N \frac{1}{\n{\v{r}_i - \v{r}_j}},
\end{equation}
where $\hat{I}_2$ is the identity in spinor space and the interaction is spinor isotropic.
We fix the number density to be $\rho$, using the same parameterization as in the standard 2DEG. This Hamiltonian is controlled by two dimensionless parameters: the familiar potential/kinetic ratio $r_s$ and a new parameter $\lambda$ that controls the concentration of Berry curvature in the lower band. We note that the single-particle part of this model can be obtained as a particular parameter choice of the two-band model discussed in 
Refs.~\cite{bernevig2006quantum}. The same Hamiltonian, up to a choice of the diagonal term, was written down in  Ref.~\cite{hu2018fractional, Tan_parent_berry}. 

\PRLsec{Single Particle Properties}Let us set out the single-particle properties of $\lambda$-jellium.
In addition to continuous translation symmetry, the Hamiltonian has $U(1)$ rotation symmetry
    $\hat{R}_\theta \phi(\v{r}) = \mathrm{diag}[1, e^{i\theta}] \phi(R_{-\theta} \v{r})$,
where the diagonal matrix acts in spinor space.
Its single-particle energy is 
$\epsilon_{\v{q}} = \frac{q^2}{r_s^2}$
in the lower band, matching the 2DEG.
The non-trivial spinor $\phi_{\v q}$ gives the lower band a skyrmionic texture in momentum space, with a spin up ``skyrmion core'' near $\v{q} = 0$ and spin down at infinity. It wraps the Bloch sphere exactly once as $\v{q}$ varies, producing Berry curvature 
$	\Omega(\v{q}) = 2 \left( \frac{\lambda}{\lambda^2 \n{\v{q}}^2 + 1} \right)^2$ (Fig.~\ref{fig:2TEG_dispersion}b)
\footnote{This non-uniform Berry curvature is in contrast to the uniform Berry curvature of the model studied in Ref.~\cite{Tan_parent_berry}.
In contrast, the $\lambda$-jellium model always has two-component spinors with Berry curvature concentrated around $\v{k}=0$.}.
Within a disk of radius $\kappa$, the enclosed Berry curvature is a Lorentzian
$	I(\kappa) = 2\pi \frac{\lambda^2 \kappa^2}{1 + \lambda^2 \kappa^2}$,
with full-width half-max $\frac{1}{\lambda}$. As $\lambda$ grows, the total Berry curvature is always $2\pi$, but becomes concentrated at $\v q=0$, reaching $I(k_F=2) = \pi$ at $\lambda = \frac{1}{2}$. The Berry curvature vanishes when $\lambda = 0$, whereupon the lower band of $\hat{H}$ reduces to the spinless 2DEG. $\lambda$-jellium therefore provides a minimal modification to the standard 2DEG whose interaction strength and band topology can be tuned continuously and independently.
The unnormalized spinors $\sqrt{1+\lambda^2 q^2}\phi_{\v q}$ are holomorphic with respect to $\v{q}$, giving the lower band of $\lambda$-jellium ``ideal quantum geometry"~\cite{JieWang_Exact_Landau_Level,Vortexability}.

\PRLsec{Mean-Field Phase Diagram}We now study the phase diagram of $\lambda$-jellium where we find a Berry-curvature induced anomalous Hall crystal competing with other crystalline phases and liquids. We work at mean-field level, employing self-consistent Hartree Fock (SCHF) with three possible translation patterns imposed: 
(I) continuous translation symmetry (CTS), (II) a discrete translation symmetry (DTS) with a triangular lattice with one electron per unit cell, (III) discrete translation symmetry (DTS) with a \textit{square} lattice with on electron per unit cell. 
The calculation with CTS is performed in the thermodynamic limit, while the DTS calculation is done on $N_k \times N_k$ unit cells, with $N_k$ going up to $36$.
See Appendices for numerical details. 
The resulting phase diagram is shown in Fig.~\ref{fig:2TEG_dispersion} (c). We now embark on a tour of the phase diagram.

\begin{figure}
    \begin{center}
    \includegraphics[width=\linewidth]{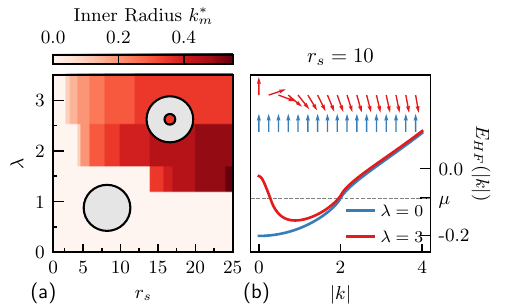}
    \end{center}
    \caption{Fermi liquid phases of $\lambda-$jellium assuming continuous translation symmetry. 
    (a) Self-consistent inner radius $k_m^*$ of the Fermi surface.
    (b) The Fermi liquid develops a strong peak at $\v k =0$ at large $\lambda,r_s$ due to exchange (Fock) interactions, leading to annular Fermi surfaces. Arrows show the pseudospin on the Bloch sphere, $(\braket{\sigma_x}, \braket{\sigma_z})$ along the $k_x$-axis. A localized ``skyrmion core" is present at $\lambda=3$.
    }
    \label{fig:FermiLiquid}
\end{figure}
\PRLsec{Circular and Annular Fermi Liquids}The phase diagram (Fig.~\ref{fig:2TEG_dispersion} (c)) supports two regions with distinct Fermi liquid ground states, whose energy competition is driven by non-uniform quantum geometry. To understand the competition between the Fermi liquids, we analyze the phase diagram while imposing continuous translation symmetry.
The mean field states are entirely characterized by their momentum space occupations $n(\v{k}) = \langle c^\dagger_{\v{k}} c_{\v{k}}\rangle$ with \textit{unbounded} momentum $\v{k}$. At small $\lambda$, we find the textbook circular Fermi liquid, with $n(\v{k}) = \theta(\n{\v{k}} - k_F)$. At moderate interaction strengths, $\lambda$ drives a transition to an annular Fermi liquid. This phase has two concentric circular Fermi surfaces, so that $n(\v{k}) = 1$ when $k_m \le \n{\v{k}} \le k_M$ and otherwise vanishes. The competition between the circular and annular Fermi liquids, which we call FL(C) and FL(A) respectively, is shown in Fig.~\ref{fig:FermiLiquid}(a).

Interactions drive the formation of the annular Fermi liquid via a mechanism that can be understood qualitatively from the skyrmionic spinor texture. This arises from the Hartree-Fock band structure~\cite{supp}
\begin{equation}
E_{\mathrm{HF}}(\v k)
= \frac{|\v{k}|^2}{r_s^2}
- \frac{1}{A}\sum_{\v{q}} V_{\v{q}} \n{\Lambda_{\v{q}}(\v{k})}^2 n(\v{k}+\v{q})
\end{equation}
where the sum runs over unrestricted momenta, $V_{\v{q}} = 2\pi/q$, and $\Lambda_{\v{q}}(\v{k}) = \phi_{\v{k}+\v{q}}^\dagger \phi_{\v{k}}$ is the form factor.
A nontrivial form factor $|\Lambda| < 1$ reduces the exchange energy gain from the Fock term, which favors ferromagnetic spinor configuration. 
How does this work with the skyrmionic spinor texture Eq.~\eqref{eq:lowerband}? For small $\lambda$, the ``skyrmion core" region where spins point up is large compared to $k_F$. The form factors are therefore close to unity, giving a large negative exchange energy (Fig~\ref{fig:FermiLiquid}b).  Conversely, large $\lambda$ spinors have a small ``skyrmion core'' compared to $k_F$, with most states inside the non-interacting Fermi surface pointing down. Occupying the mis-aligned spinors near the core incurs a significant energy penalty.
This drives the state to deplete the skyrmion core, resulting in an annular Fermi surface, Fig.~\ref{fig:FermiLiquid}(a). The depletion of $\v{k}=0$ will be a recurring feature throughout the phase diagram.

\PRLsec{Halo Wigner Crystals}Wigner crystal phases are characterized by spontaneous breaking of continuous symmetry that produces an insulator with zero Chern number.
The 2DEG is known to host triangular Wigner crystalline states at large $r_s$~\cite{Bonsall-Maradudin,Meissner_Stability_and,RMP_2D_electron}. Mean-field calculations underestimate the value of critical $r_s$ between ferromagnetic FL and WC to be around $2$~\cite{HF_Wigner_crystal}. The $\lambda$-jellium model phase diagram, which is continuously connected to 2DEG in the $\lambda\to 0$ limit, also hosts Wigner crystalline phases, shown in Fig.~\ref{fig:2TEG_dispersion}(c).

At large $r_s$ and $\lambda$, there is a distinct Wigner crystal phase --- also with a triangular lattice --- which we call a halo Wigner crystal. A similar crystalline phase was recently studied in Bernal bilayer graphene~\cite{joy2023wignercrystallizationbernalbilayer}. To see this,    we plot the momentum space occupation number $n(\v k) = \langle c^\dagger_{\v{k}} c_{\v{k}} \rangle$ for a normal Wigner crystal and a halo Wigner crystal in Fig.~\ref{fig:WC}(a) and (b). We see that $n(\v k=0)$ takes the maximal value for the normal Wigner crystal, while $n(\v k=0) = 0$ for the halo Wigner crystal.

The vanishing of $n(\v k=0)$ is in fact enforced by how the crystal transforms under $C_3$ symmetry.
Thus, the orbitals occupied at the high symmetry points must be eigenstates of the $C_3$ symmetry with a definite angular momentum.
Concretely, the $C_3$ angular momentum is
\begin{equation}
    \hat{C}_3 \Psi_{\Gamma}^{\mathrm{WC}} =  \Psi_{\Gamma}^{\mathrm{WC}} , \quad
    \hat{C}_3 \Psi_{\Gamma}^{\mathrm{halo WC}} = e^{i 2\pi/3} \Psi_{\Gamma}^{\mathrm{halo WC}},
\end{equation}
where $\Psi_\Gamma$ is the single-particle orbital at the $\Gamma$ point.
Since the angular momentum at $\Gamma$ does not depend on the choice of $C_3$ centers, the halo Wigner crystal and the normal Wigner crystal are distinct crystalline insulators with different symmetry properties~\cite{po2020symmetry}. This transition can in fact be undersood from a simple semiclassical analysis~\cite{joy2023wignercrystallizationbernalbilayer}, as we review in~\cite{supp}.

\begin{figure}
    \centering
    \includegraphics[width=\linewidth]{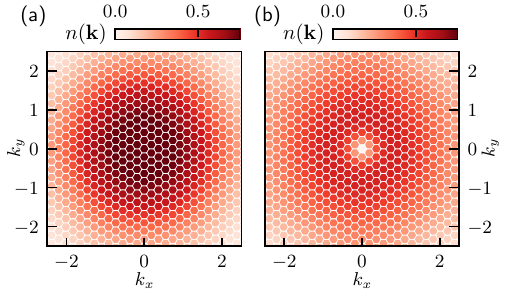}
    \caption{Occupations of plane wave states in the lower band for different Wigner crystals. (a) Normal Wigner crystal at $(r_s,\lambda) = (20,0)$. (b) Halo Wigner crystal at $(r_s,\lambda) = (20,2.5)$. The $\v k=0$ region is depleted due to the Fock interactions at large $\lambda$. %The black lines corresponds to the boundary of the first Brillouin zone.
        }
    \label{fig:WC}
\end{figure}

\PRLsec{Anomalous Hall Crystal} 
Due to the coexistence of single-particle Berry curvature and a trend towards crystallization at large interactions, we expect the anomalous Hall crystal (AHC) --- a crystalline phase that becomes a Chern insulator when pinned --- to appear~\cite{Zeng_sublattice_structure,AHC2,Zhihuan_Stability,Tan_parent_berry,Tan_FAHC}. 
At moderate $r_s$, we indeed observe an AHC phase. The phase boundary from WC to AHC occurs around $\lambda=1/2$ where the Berry curvature within the Fermi surface reaches $\pi$ --- consistent with the prediction of Ref.~\cite{Zhihuan_Stability}. At large $r_s$, the interaction mixes the single-particles bands strongly, and the AHC is overtaken by the halo WC.

At even larger $\lambda \gtrsim 2.5$, the AHC undergoes a transition to the annular Fermi liquid.
Finite size effects are particularly acute in the large $\lambda$ regime due to the tiny skyrmion core in momentum space.

\begin{figure}[t]
    \centering
    \includegraphics[width=\linewidth]{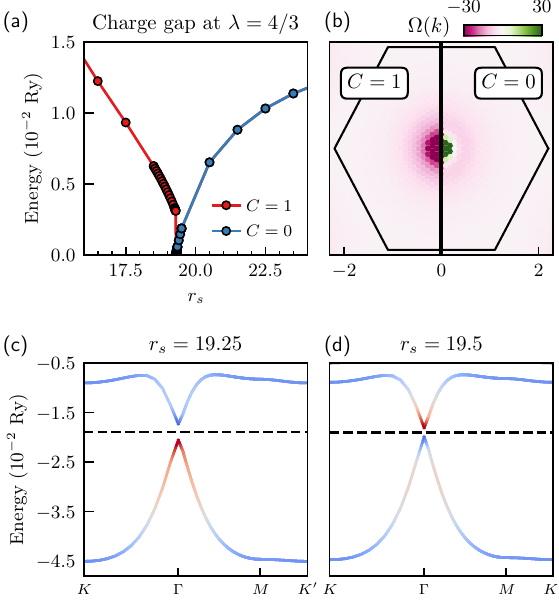}
    \caption{Continuous phase transition driven by $r_s$ at $\lambda=4/3$.
    (a) Direct charge gap near the phase boundary between AHC and the halo WC.
    (b) The Berry curvature of an AHC ($r_s = 19.25$) and WC ($r_s = 19.5$) near the transition. It is strongly peaked around the $\Gamma$ point, saturating the color scale.
    (c,d) self-consistent Hartree-Fock band structures before and after the transitions. Coloring corresponds to $z_a(\v p)$, with red corresponding to occupying the first component. Clearly, the $r_s$ tuned transition corresponds to a band inversion transition. 
    }
    \label{fig:PT}
\end{figure}

\PRLsec{Second-order topological phase transition}We now examine the crystal-to-crystal phase transitions in $\lambda$-jellium, shown in Fig.~\ref{fig:2TEG_dispersion}(c). The transitions between AHC to WC, and WC to halo-WC are first-order, as the charge gap remains finite across the transitions.
The transition between halo WC and AHC, on the other hand, appears to be continuous, as the charge gap closes at the transition(Fig.~\ref{fig:PT}(a)).

The origin of this continuous transition can be understood from the Hartree-Fock band structure in Fig.~\ref{fig:PT}(c,d). There the bands are colored according to their spinor polarization $z_a(\v k)=\braket{\psi_{a\v p}|\sigma_z|\psi_{a\v p}}$, where $\sigma_z$ acts in spinor space and $\ket{\psi_{a\v p}}$ is the Bloch state at $\v{p}$ for band $a=1,2$, counting from below. The spinor polarization flips at the transition, signaling change in angular momentum. This is accompanied by a change in the sign of Berry curvature around the $\Gamma$ point (Fig.~\ref{fig:PT}(b)). 
The vicinity of $\Gamma$ is therefore well-described by a massive Dirac model with mass inversion.

\PRLsec{AHC Lattice Geometry}In the \textit{classical} Wigner crystal $r_s \to \infty$,
the triangular lattice is preferred over the square lattice by about $0.5\%$~\cite{Bonsall-Maradudin}. A square lattice, however, can be stabilized at small $r_s$~\cite{HF_Wigner_crystal}. The preference for a triangular lattice therefore depends on the specific details of the Wigner crystal phase.

Does the AHC prefer a triangular lattice? To understand this, we examine the energetic competition within the AHC phase between the triangular lattice and the square lattice. Our result, shown in Fig.~\ref{fig:2TEG_dispersion}(c), is that while the triangular lattice is preferred over the square in most of the AHC phase diagram, the reverse is true near the phase boundary between normal WC and AHC. The energetic competition remains close, with the square lattice favored by $<0.5\%$ at most for $\lambda = 2/3$.
In contrast, the triangular lattice is more stable at larger $\lambda$ and $r_s$. We caution that another unit cell shape may have yet-lower energy in this region; future work will examine the full landscape of possible unit cells~\cite{AHC4}.

\PRLsec{Discussion}This work studied $\lambda$-jellium, a minimal extension of jellium with topological crystalline phases including an anomalous Hall crystal. Its phase diagram 
 captures the key features of more complex microscopic models \cite{AHC_Senthil,AHC_Yahui,AHC1,guo2024fractional,kwan2023moir,Tan_parent_berry} at mean field level, suggesting universality in topological band minima.

A crucial next step is to move beyond mean-field techniques. As mentioned above, the two-component nature of our model together with the fact that only first and second derivatives appear make it well-suited for a variety of many-body numerical techniques, including variational monte carlo and neural network wavefunction methods. $\lambda$-jellium may therefore be a good model to establish crystallization with non-trivial Chern number at the many-body level. 

\begin{acknowledgements}
We thank Ophelia Evelyn Sommer, Taige Wang, Tianle Wang, Mike Zaletel, Patrick Ledwith and Eslam Khalaf  for related collaborations and useful insights.
We acknowledge Erez Berg, Yaar Vituri, Agnes Valenti, Miguel Morales, Paul Yang,  Shiwei Zhang, F\'{e}lix Desrochers,  Yong Baek Kim, Adrian Po and Trithep Devakul for fruitful discussions.
This research was supported in part by grant NSF PHY-2309135 to the Kavli Institute for Theoretical Physics (KITP).
This research is funded in part by the
Gordon and Betty Moore Foundation’s EPiQS Initiative,
Grant GBMF8683 to T.S.   A.V. and J.D. were funded by NSF DMR-2220703. DEP acknowledges startup funds from UC San Diego. This work used the Expanse cluster at the San Diego Supercomputer Center through allocation PHY240272 from the Advanced Cyberinfrastructure Coordination Ecosystem: Services \& Support (ACCESS) program, which is supported by U.S. National Science Foundation grants No.2138259, No.2138286, No.2138307, No.2137603, and No.2138296.
\end{acknowledgements}

\bibliography{references}
\bibliographystyle{unsrt}

\onecolumngrid
\appendix
\newpage

\begin{center}
\textbf{\large Appendix}
\end{center}% \documentclass[prb,aps,onecolumn,longbibliography,superscriptaddress,preprintnumbers]{revtex4-2}

\title{
Supplemental material for: A Jellium Model for the Anomalous Hall Crystal
}

\author{Tomohiro Soejima (\begin{CJK*}{UTF8}{bsmi}副島智大\end{CJK*})}
\thanks{These authors contributed equally.}
\affiliation{Department of Physics, Harvard University, Cambridge, MA 02138, USA}

\author{Junkai Dong (\begin{CJK*}{UTF8}{bsmi}董焌\end{CJK*}\begin{CJK*}{UTF8}{gbsn}锴\end{CJK*})}
\thanks{These authors contributed equally.}
\affiliation{Department of Physics, Harvard University, Cambridge, MA 02138, USA}
\affiliation{Kavli Institute for Theoretical Physics, University of California, Santa Barbara, California 93106, USA}

\author{Ashvin Vishwanath}
\affiliation{Department of Physics, Harvard University, Cambridge, MA 02138, USA}

\author{Daniel E. Parker}
\affiliation{Department of Physics, University of California at San Diego, La Jolla, California 92093, USA}

\maketitle

\tableofcontents

\section{Construction of $\lambda$-Jellium}

\begin{figure}
    \centering
    \includegraphics[width=0.6\linewidth]{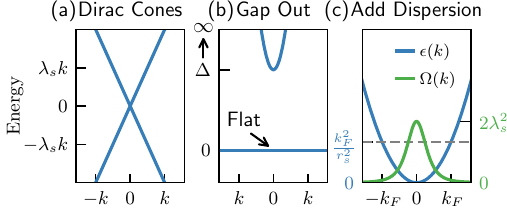}
    \caption{Construction of the $\lambda$-jellium dispersion. (a) Dirac Hamiltonian. (b) Adding a momentum-dependent mass makes the lower band exactly flat with Berry curvature. (c) Adding an spinor-isotropic kinetic term gives a quadratic dispersion with an independently adjustable Berry curvature distribution.}
    \label{fig:lambda_jellium_construction}
\end{figure}

This Appendix constructs and physically motivates the kinetic term of $\lambda$-jellium in Eq.~\eqref{eq:holomorphic_flat_band}. We start with a Dirac Hamiltonian with velocity $\lambda$ (Fig~\ref{fig:lambda_jellium_construction}a). To create Berry curvature, we must gap out the Dirac point. Ordinarily this would create a concave down dispersion, unlike jellium, which we eliminate using an momentum dependent mass. We thus choose
\begin{equation}
    \hat{h}_0 = \Delta \begin{bmatrix}
        0 & -i\lambda_s \partial\\
        -i\lambda_s \bar{\partial} & 0
    \end{bmatrix}
    +
    \Delta
    \begin{bmatrix}
        -\lambda_s^2 \nabla^2 & 0\\
        0 & 1
    \end{bmatrix},
\end{equation}
so that the lowest band dispersion becomes $\epsilon_{\v{q}} = 0$, i.e. it is exactly flat (Fig~\ref{fig:lambda_jellium_construction}b). The constant $\Delta$ sets the minimum gap to the upper band. The spinor of the lower band wavefunction is given by Eq.~\eqref{eq:lowerband} in the main text, and $\lambda$ controls its Berry curvature distribution. As the wavefunction of the lower band is independent of $\Delta$, we may take $\Delta \to \infty$ and restrict our attention to the lower band. Finally, we add the kinetic term of normal jellium:
\begin{equation}
    \hat{h} = \hat{h}_0 -\frac{\nabla^2}{r_s^2} \begin{bmatrix}
        1 & 0\\
        0 & 1
    \end{bmatrix}.
\end{equation}
This modifies the dispersion of the lower band to $\epsilon_{\v{q}} = \n{q}^2/r_s^2$, matching normal jellium, while leaving its single particle wavefunction Eq.~\eqref{eq:lowerband} unchanged. 
(Fig~\ref{fig:lambda_jellium_construction}c).

\section{Modeling Details}
\label{app:model_details}

This appendix describes modeling details of the two-dimensional topological electron gas.

\subsection{Geometry}
We consider electrons in the infinite plane subject to the Hamiltonian in Eq.~\eqref{eq:ham} with a compensating background charge. To make the computation tractable, we impose a periodic boundary condition, and consider a Hilbert space defined by a finite torus $M$ with area $A = N\rho$, where $N$ is the number of particles. When imposing discrete translation symmetry, we choose $M$ to be compatible with the choice of the Bravais lattice.

\subsubsection{Triangular lattice}

Consider a direct lattice $\mathbb{L}$ with lattice vectors
\begin{equation}
    \v{a}_1 = L \begin{pmatrix}
    \frac{1}{2}, & 
    -\frac{\sqrt{3}}{2}
    \end{pmatrix}, \quad
     \v{a}_2 = L \begin{pmatrix}
    \frac{1}{2}, &
    \frac{\sqrt{3}}{2}
    \end{pmatrix}.
\end{equation}
Here $L = \sqrt{\frac{2\pi}{\sqrt{3}}} a$ ensures the density $\rho = \frac{1}{\pi a^2}$ matches the unit cell size $\Auc = L^2 \frac{\sqrt{3}}{2}$. The reciprocal lattice is $\mathbb{L}^*$ with reciprocal basis vectors
\begin{equation}
    \v{g}_1 = \frac{2\pi}{L} \begin{pmatrix}
    1, & 
    -\frac{1}{\sqrt{3}}
    \end{pmatrix}, \quad
     \v{g}_2 = \frac{2\pi}{L} \begin{pmatrix}
    1, &
    \frac{1}{\sqrt{3}}
    \end{pmatrix}.
\end{equation}
The full torus is a parallelogram with $N = N_1 \times N_2$ unit cells and total area $A = N A_{\text{uc}}$ with $\Auc = \pi a^2$.

\subsubsection{Square lattice}
The direct lattice is given by
\begin{equation}
    \v{a}_1 = L \begin{pmatrix}
    1, & 
    0
    \end{pmatrix}, \quad
     \v{a}_2 = L \begin{pmatrix}
    0, &
    1
    \end{pmatrix}.
\end{equation}
Here we take $L = \sqrt{\pi} a$. The reciprocal lattice is $\mathbb{L}^*$ with reciprocal basis vectors
\begin{equation}
    \v{g}_1 = \frac{2\pi}{L} \begin{pmatrix}
    1, & 
    0
    \end{pmatrix}, \quad
     \v{g}_2 = \frac{2\pi}{L} \begin{pmatrix}
    0, &
    1
    \end{pmatrix}.
\end{equation}
The full torus is a parallelogram with $N = N_1 \times N_2$ unit cells and total area $A = N A_{\text{uc}}$ with $\Auc = \pi a^2$.

\subsubsection{Brillouin Zone \& Momentum Grid}

We consider a parallelogram Brillouin zone 
\begin{equation}
	\bz = \setc{k_1 \v{g}_1 + k_2 \v{g}_2}{-\frac{1}{2} \le k_{1,2} < \frac{1}{2}}.
\end{equation}
We denote elements of the Brillouin zone by $\v{k} \in \bz$, and unrestricted momenta by $\v{q}$. Such momenta can always be split as
\begin{equation}
	\v{q} = \{\v{q}\}  + [\v{q}] = \v{k} + \v{g},
\end{equation}
where $\v{k}$ is in the first Brillouin zone and $\v{g} = n_1 \v{g}_1 + n_2 \v{g}_2$ is a reciprocal lattice vector.

We take a Monkhorst-Pack discretization of the Brillouin zone
\begin{equation}
\v{k}_{n_1,n_2} = \frac{n_1 + \Phi_1/2\pi}{N_1} \v{g}_1
+ \frac{n_2 + \Phi_2/2\pi}{N_2} \v{g}_2,
\end{equation}
where $0 \le \Phi_{1,2} \le 2\pi$ are the two fluxes through the torus and
\begin{equation}
    n_i = \begin{cases}
-\frac{N_i}{2}, -\frac{N_i}{2} + 1, \dots, \frac{N_i}{2} -1
    & \text{ for $N_i$ even}\\
-\frac{N_i}{2} + \frac{1}{2}, -\frac{N_i}{2} + \frac{3}{2}, \dots, \frac{N_i}{2} - \frac{1}{2}
    & \text{ for $N_i$ odd.}
    \end{cases}
\end{equation}
This choice ensures that the lattice goes through $\Gamma$ for $\Phi_{1,2} = 0$.

\subsubsection{Momentum Shell Structure}

To describe continuous functions within the unit cell (such as wavefunctions), we consider a ``shell structure" of reciprocal momenta
\begin{equation}
	S = \setc{\v{g} = m_1 \v{g}_1 + m_2 \v{g}_2}{ \n{\v{g}} < N_s \max(\n{\v{g}_1},\n{\v{g}_2})}.
\end{equation}
In practice we usually take $N_s = 5-7$, giving $\n{S} \approx 100-200$.

\subsection{First-Quantized Wavefunctions}

We now set out basis conventions for first-quantized wavefunctions.

\subsubsection{Microscopic Basis}

Consider a microscopic basis $\ket{\v{r}, I}$ where $I = 1,2$ is a spinor index with corresponding basis vectors $\ev_I$. Wavefunctions are then written as
\begin{equation}
	\psi^I(\v{r}) = \braket{\v{r},I|\psi}.
\end{equation}
Spinor indices are usually left implicit for concision.

\subsubsection{Bloch Theorem Conventions}

Suppose $\hat{h}$ has discrete translation symmetry: $[\hat{h},\hat{T}_{\v{R}}] = 0$ where $\hat{T}_{\v{R}} f(\v{r}) = f(\v{r}+\v{R})$ is translation by a direct lattice vector. We then consider Bloch wavefunctions
\begin{align*}
	\hat{h} \ket{\psi_{\v{k} n}} &= \varepsilon_{\v{k} n} \ket{\psi_{\v{k}n}}\\
	\hat{T}_{\v{R}} \ket{\psi_{\v{k} n}} &= e^{i\v{k}\cdot\v{r}} \ket{\psi_{\v{k}n}}\\
	\phi^I_{\v{k}n}(\v{r}) &= \braket{\v{r},I|\psi_{\v{k}n}}.
\end{align*}
Here $n$ labels bands. The periodic part of the Bloch waves are defined as
\begin{equation}
	u^I_{\v{k} n}(\v{r}) = N^{\frac{1}{2}} e^{-i\v{k}\cdot\v{r}} \phi_{\v{k}n}^I(\v{r}),
\end{equation}
normalized on the unit cell. These are eigenstates of $\hat{h}[\v{k}] = e^{-i\v{k}\cdot \hat{\v{r}}} \hat{h} e^{i\v{k}\cdot \hat{\v{r}}}$. The Bloch states comprise a unitary transform:
\begin{align}
	\braket{\psi_{\v{k}n}|\psi_{\v{k}'n'}} &= 
	\sum_{I} \int_M d^2 \v{r}\; \overline{\phi^I_{\v{k}n}(\v{r})} \phi_{\v{k}'n'}^I(\v{r}) = \delta_{\v{k}\v{k}'} \delta_{nn'},\\
	\braket{\v{r},I|\v{r}',I'} &= \sum_n \sum_{\v{k} \in \bz} \overline{\phi^I_{\v{k}n}(\v{r})} \phi_{\v{k}n}^{I'}(\v{r}') = \delta^{II'} \delta(\v{r}-\v{r}').
\end{align}
Note that the $\v{k}$ sum is discrete due to the finite area $A$.

We adopt a periodic gauge so that
\begin{equation}
	\ket{\psi_{\v{k}+\v{g},n}} = \ket{\psi_{\v{k},n}}.
\end{equation}

% We adopt a periodic gauge so that
% \begin{equation}
% 	\ket{\psi_{\v{k}+\v{g},n}} = \hat{V}_{\v{g}} \ket{\psi_{\v{k},n}}, \quad \hat{V}_{\v{g}} = e^{-i\v{g}\cdot\hat{\v{r}}}.
% \end{equation}
% Though this operator is unitary in principle, in practice there is a parametrically small violation $\hat{V}_{\v{g}} \hat{V}_{\v{g}}^\dagger = I + O(N_s)$.

\subsubsection{Computational Basis}

The computational basis is in momentum space, with unbounded momentum $\v{q}$ partitioned as $\v{k}+\v{g}$ as above. Explicitly, we use
\begin{equation}
	\braket{\v{r},I|\v{k},\v{g},I'} = \frac{1}{\sqrt{A}} e^{i(\v{k}+\v{g})\cdot\v{r}} \delta_{II'}
\end{equation}
where $\v{g} = m_1 \v{g}_1 + m_2\v{g}_2$ and $\v{k} \in \bz$.  The discretization of $\v{k}$ is given above, as is the cutoff on $\v{g}$.

We use the following Fourier conventions:
\begin{align}
	\phi_{\v{k} n}^I(\v{r}) &= \sum_{\v{k}',\v{g},I'} \braket{\v{r},I|\v{k}',\v{g},I'}\braket{\v{k}',\v{g},I'|\psi_{\v{k},n}}\\
							&= \frac{1}{\sqrt{A}} \sum_{\v{g}}  \, e^{i (\v{k}+\v{g})\cdot\v{r}} \phi^{\v{g} I}_{\v{k} n},\\
	\phi^{\v{g},I}_{\v{k},n} &=  \braket{\v{k},\v{g},I|\psi_{\v{k},n}} = \sum_I \int_M d^2 \v{r}\;  \frac{e^{-i(\v{k}+\v{g})\cdot\v{r}}}{\sqrt{A}} \phi_{\v{k},n}^I(\v{r}).
\end{align}
The computational eigenvectors are thus normalized as 
\begin{equation}
	\sum_{\v{g},I} \overline{\phi^{\v{g},I}_{\v{k},n}} \phi_{\v{k}',n'}^{\v{g},I} = \delta_{nn'}.
\end{equation}

\subsection{Interactions}

We consider Coulomb interactions between electrons
\begin{align}
	V(\v{r}) &=  \frac{2}{r_s\n{\v{r}}} = \frac{1}{A}\sum_{\v q} V(\v q) e^{i \v{q} \cdot \v{r}},\\
	V(\v q)  &= \frac{4\pi}{r_s q}.
\end{align}
This Hamiltonian suffers from IR divergence due to the long-range Coulomb interaction, and thus needs to be accompanied by a neutralizing background charge. We review the procedure for correcting for neutralizing background charge in App.~\ref{app:ewald}. The resulting interaction on the torus takes the form

\begin{equation}
    H_\mathrm{int} = \frac{1}{2A} \sum_{\v{q}\neq 0} V(\v q) :\hat{\rho}_{\v{q}} \hat{\rho}_{-\v{q}}: + \Delta \epsilon,
\end{equation}
where $:\cdots :$ is the normal ordering operator with respect to the vacuum. We remove $\v{q} = 0$ from the sum due to cancellation with the background charge, and $\Delta \epsilon$ encodes the energy correction from self-images. Formally, it is written as

\begin{align}
    \Delta \epsilon
    =
    \frac{1}{N}
    \left(
        -\frac{\rho}{2} V_{\v q=0}
        + \frac{1}{2} N \sum_{\v{R} \neq 0} V(\v{R})
    \right),
\end{align}
where $\v{R}$ is the set of vectors that define the periodic boundary condition of the torus $M$. Although both of the terms are formally divergent, the divergences cancel each other. In practice, we need to perform an Ewald summation to compute this summation. Including the correction $\Delta \epsilon$ handles the dominant contribution to finite-size energy correction of the system (see Fig.~\ref{eq:finite_size_Madelung} below from the last Appendix).

\subsection{Second quantization and band projection}
We now perform second quantization and band projection of the Hamiltonian. Let us denote by $c^\dagger_{\v{k}}$ the creation operator for the lower band. The kinetic part of the Hamiltonian is simply given by

\begin{equation}
    H_{\mathrm{kin}} = \sum_{\v{k}} \frac{|\v{k}|^2}{r_s^2} c^\dagger_{\v{k}} c_{\v{k}}
\end{equation}

The non-constant part of the interaction is given by

\begin{equation}
    \frac{1}{2A} \sum_{\v{q}\neq 0} V(\v q) :\hat{\rho}_{\v{q}} \hat{\rho}_{-\v{q}}:
\end{equation}

We can write the band projected interaction as 
\begin{equation}
    \tilde{\rho}_{\v{q}} = \sum_{\v{k}} \Lambda_{\v q}(\v k) c^\dagger_{\v{k}} c_{\v{k} + \v{q}}
\end{equation}
where $\Lambda_{\v q}(\v k) = \braket{u_{\v{k}} | u_{\v{k}+\v{q}}}$.
The projected interaction is therefore written as

\begin{align}
    H_\mathrm{int} &= \frac{1}{2A} \sum_{\v{q}\neq 0} \sum_{\v{k}, \v{k}'} V(\v{q}) \Lambda_{\v{q}}(\v{k}) \Lambda_{-\v{q}}(\v{k}'):c^\dagger_{\v{k}} c_{\v{k}+\v{q}} c^\dagger_{\v{k}'} c_{\v{k}' - \v{q}}:.
\end{align}

\section{Review of Ewald summation and Madelung energy}
\label{app:ewald}

We now review the method of Ewald summation and computation of the Madelung energy. Some useful references are found in Ref.~\cite{kawata2001rapid, parry1975electrostatic}.

We consider a system of electrons in the infinite plane. We denote their interaction by $V(\v{r})$. The interaction energy is given by a formal sum

\begin{equation}
    E_\mathrm{{int}} = \sum_{i < j}^\infty V(\v{r}_i - \v{r}_j).
\end{equation}

Our goal is to evaluate the energy density of this system by imposing periodic boundary condition.
We consider a parallelogram $P = \v{R}_1 \wedge \v{R}_2$ spanned by $\v{R}_1$ and $\v{R}_2$. This defines the lattice $\Lambda = \{m\v{R}_1 + n\v{R}_2 | m,n \in \mathbb{Z} \}$. It is convenient to write $\v{R}_j = L_j \v{r}_j$ with $L_i \in \mathbb{Z}$. We take $|\hat{z} \cdot \v{r}_1 \times \v{r}_2| = 1/\rho$, where $\rho$ is the electron density. The parallelogram therefore contains $N = L_1L_2$ electrons.

We can now truncate the formal sum to define energy density. To do so, let us rewrite the formal sum as follows:

\begin{align}
E_{\mathrm{int}} &= \sum_{i < j}^\infty V(\v r_i - \v r_j)
\\
&= \frac{1}{2}\sum_{i \neq j}^\infty V(\v r_i - 
\v r_j)
\\
&= \frac{1}{2}\sum_{i \neq j}^N \sum_{\v R_a, \v R_b \in \Lambda} V(\v r_i -\v R_a - \v r_j - \v R_b) + \frac{1}{2} \sum_i^N\sum_{\v R_a \neq \v R_b} V(\v R_a - \v R_b)
\\
&= \sum_{\v R_a \in \Lambda} \frac{1}{2} \sum_{i \neq j}^N \sum_{\v R_b \in \Lambda} V(\v r_i - \v r_j - \v R_b|) + \sum_{\v R_a} \frac{1}{2}\sum_i^N \sum_{\v R_b \neq 0} V(\v R_b)
\\
&=\sum_{\v R_a}
\left(
\sum_{i < j}^N \sum_{\v R_b \in \Lambda} V(\v r_i - \v r_j - \v R_b|) + \frac{1}{2} \sum_i^N \sum_{\v R_b \neq 0} V(\v R_b).
\right)
\end{align}
In going from the second line to the third line, we introduced the sum over positions within the unit cell i.e. $\sum_i^\infty f(\v r_i)$ is the sum over all particle positions, and $\sum_i^N f(\v r_i)$ is a sum over particle positions within the unit cell.
We can split the sum to the part $i\neq j$, in which case there is no restriction on values of $\v R_a$ and $\v R_b$, and $i=j$, in which case we require $\v R_a \neq \v R_b$.
The energy par particle is then given by 
% \TS{I am now fairly sure about the factor of $\frac{1}{2}$}:

\begin{equation}
    \epsilon_{\mathrm{int}} = \frac{1}{N}
    \left(
        \sum_{i<j}^{N} \sum_{\v{R} \in \Lambda } V(\v{r}_i - \v{r}_j - \v{R})
        + \frac{1}{2}\sum_{i}^{N} \sum_{\v{R} \neq \v{0}} V(\v{R})
    \right),
\end{equation}
where $\v{r}_i$'s label the location of $N$ electrons within the parallelogram.

If $V(\v r)$ is the Coulomb interaction, this sum is infrared divergent. We can regularize this divergence by adding a neutralizing background:

\begin{align}
    \epsilon_{\mathrm{bkg}} &=
    \frac{1}{N}
    \left(
    -\sum_{i}^N \int \rho_b(\v r) V(\v r) d^2r + \frac{1}{2}\int_{P} d^2 \v r \int_{\mathbb{R}^2} d^2 \v r' \rho_b(\v r) \rho_b(\v r') V(\v r-\v r')
    \right)
    \\
    &=
    \frac{1}{N}
    \left(
    -N \rho \int V(\v r) d^2\v r + \frac{1}{2}\frac{N}{\rho} \rho^2 \int d^2 \v r V(\v r)
    \right)
    \\
    &=
   - \rho \int V(\v r) d^2\v r + \frac{1}{2} \int d^2 \v r V(\v r) = -\frac{\rho}{2} \int d^2 \v r  V(\v r)
\end{align}
where we have chosen the neutralizing background to be uniform and exactly cancel the charges of the electrons: $\rho_b(\v r)=\rho$. In the sum, the first term is the electron-background interaction, and the second term is the background-background interaction.

We now wish to evaluate this expression in momentum space. To do so, we first assume $V(r)$ is properly regularized such that $V(\v k=0) = \int d^2 \v r V(\v r)$ is finite. The precise regularization scheme does not matter.
The interaction term can then be rearranged as 

\begin{align}
    \epsilon_{\mathrm{int}}
    &= \frac{1}{N}
    \left(
        \sum_{i<j}^{N} \sum_{\v{R} \in \Lambda } V(\v{r}_i - \v{r}_j - \v{R})
        + \frac{1}{2}\sum_{i}^{N} \sum_{\v{R} \neq 0} V(\v{R})
    \right)
    \\
    &=
    \frac{1}{N}
    \left(
        \frac{1 }{|P|}
        \sum_{i<j}^{N}
        \sum_{\v{G} \in \Lambda^*} \tilde{V}(\v{G})e^{i \v{G} \cdot (\v{r}_i - \v{r}_j)}
        + \frac{1}{2}\sum_{i}^{N} \sum_{\v{R} \neq 0} V(\v{R})
    \right)
    \\
    &=
    \frac{1}{N}
    \left(
        \frac{(N - 1)\rho}{2} \tilde{V}(0)
        + \frac{\rho}{N} \sum_{i<j}^{N}\sum_{\v{G} \neq \v{0}}  \tilde{V}(\v{G})e^{i \v{G} \cdot (\v{r}_i - \v{r}_j)}
                + \frac{1}{2} \sum_{i}^{N} \sum_{\v{R} \neq 0} V(\v{R})
    \right)
\end{align}
where we used Poisson summation: $\sum_{\v{R} \in \Lambda} f(\v{R}) = \frac{1}{|P|}\sum_{\v{G} \in \Lambda^*} \tilde{f}(\v{G}), |P|=N/\rho$. From now on, we will switch to the notation where $\tilde{f}$ is the Fourier transform of $f$. 
The neutralizing background, on the other hand, gives

\begin{equation}
    \epsilon_{\mathrm{bkg}} = -\frac{\rho}{2} \int d^2 \v r  V(\v r) = -\frac{1}{2}\rho \tilde{V}(0). 
\end{equation}

Combining these together, we get

\begin{align}
    \epsilon_{\mathrm{int}} + \epsilon_{\mathrm{bkg}}
    =
    \frac{1}{N}
    \left(
        -\frac{\rho}{2} \tilde{V}(0)
        +
                \frac{ \rho}{N}\sum_{i<j}^{N}\sum_{\v{G} \neq \v{0}} \tilde{V}(\v{G})e^{i \v{G} \cdot (\v{r}_i - \v{r}_j)}
    + \frac{1}{2}\sum_{i}^{N} \sum_{\v{R} \neq 0} V(\v{R})
    \right)
\end{align}

The first term and the last term  (Madelung term) do not scale with the number of particles $N$, so they can be dropped to give us
\begin{equation}
    \epsilon_{\mathrm{int}} + \epsilon_{\mathrm{bkg}} 
    =
                \frac{\rho}{N^2}\sum_{i < j}^{N}\sum_{\v{G} < \v{0}} \tilde{V}(\v{G})e^{i \v{G} \cdot (\v{r}_i - \v{r}_j)} + O\left(\frac{1}{N}\right).
\end{equation}

We now treat the $O(1/N)$ term carefully to tame the finite-size effect. The correction term is

\begin{align}
    \Delta \epsilon
    =
    \frac{1}{N}
    \left(
        -\frac{\rho}{2} \tilde{V}(0)
        + \frac{1}{2} N \sum_{\v{R} \neq 0} V(\v{R})
    \right).
\end{align}

We will now use the Ewald summation trick and split the interaction as $V(r) = V^\ell(r) + V^s(r)$. We demand $V^\ell(0) < \infty$ (UV convergent, IR divergent) and $\tilde{V}^s(k=0) < \infty$ (UV divergent, IR convergent). 
We can now rewrite the sum as

\begin{align}
    \Delta \epsilon
    &=
    \frac{1}{N}
    \left(
        -\frac{ \rho}{2} \tilde{V}^\ell(0)
        + \frac{1}{2} N \sum_{\v{R} \neq 0} V^\ell(\v{R})
        -\frac{ \rho}{2} \tilde{V}^s(0)
        + \frac{1}{2} N \sum_{\v{R} \neq 0} V^s(\v{R})
    \right)
    \\
    &=
    \frac{1}{N}
    \left(
        -\frac{ \rho}{2} \tilde{V}^\ell(0)
        + \frac{1}{2} N \sum_{\v{R}} V^\ell(\v{R}) - \frac{N}{2}V^\ell(0)
        -\frac{ \rho}{2} \tilde{V}^s(0)
        + \frac{1}{2} N \sum_{\v{R} \neq 0} V^s(\v{R})
    \right)
    \\
    &=
    \frac{1}{N}
    \left(
        -\frac{ \rho}{2} \tilde{V}^\ell(0)
        + \frac{  \rho}{2} \sum_{\v{G}} \tilde{V}^\ell(\v{G}) - \frac{N}{2} V^\ell(0)
        -\frac{ \rho}{2} \tilde{V}^s(0)
        + \frac{1}{2} N \sum_{\v{R} \neq 0} V^s(\v{R})
    \right)
    \\
    &= 
    \frac{1}{N}
    \left(
        \frac{  \rho}{2} \sum_{\v{G}\neq 0} \tilde{V}^\ell(\v{G}) - \frac{N}{2}V^\ell(0)
        -\frac{ \rho}{2} \tilde{V}^s(0)
        + \frac{1}{2} N \sum_{\v{R} \neq 0} V^s(\v{R})
    \right)
\end{align}

All terms in this expression are manifestly finite, and we can evaluate them directly.

\subsection{Error function for the Ewald trick}
We now make a choice for decomposition $V(\v r) = V(|\v r|) = V^\ell(|\v r|) + V^s(|\v r|)$. The standard method is to take the following:

\begin{equation}
    V^s(r) = \frac{\mathrm{erfc}(\alpha r)}{r}, \quad V^\ell(r) = \frac{\mathrm{erf}(\alpha r)}{r},
\end{equation}
where $\mathrm{erf}(r) = \frac{2}{\sqrt{\pi}}\int_0^r e^{-t^2} dt, \mathrm{erfc}(r) = 1-\mathrm{erf}(r)$. There is a simple physical picture for this choice. $V^\ell(r)$ is the potential coming from a gaussian cloud of electron with charge density $e^{-\alpha r^2}$, as can be checked from the Gauss law.

% \begin{figure}
%     \centering
%     \includegraphics[width=0.6\linewidth]{Notes/ewald_comparison.pdf}
%     \caption{Comparison of the behavior of short and long range part of Ewald terms at $\alpha=1$.}
%     \label{fig:enter-label}
% \end{figure}

\subsection{3D Ewald}
Even though we are interested in 2D, it is convenient to start from evaluating the 3D integral first. The 3D Fourier transform of $V^\ell(r)$ is

\begin{equation}
    \tilde{V}^\ell(k) = 4\pi \frac{e^{-\frac{k^2}{4\alpha^2}}}{k^2}
\end{equation}

We check this by performing a Fourier transform:

\begin{align}
    \mathcal{F}[\tilde{V}^\ell](r) &= 4\pi\int \frac{d^3 k}{(2\pi)^3} \frac{e^{-\frac{k^2}{4\alpha^2}}}{k^2} e^{i k \cdot r}
    \\
    &= \frac{1}{\pi}\int_0^{\infty} dk e^{-\frac{k^2}{4\alpha^2}} \int \sin \theta d\theta e^{i k |r| \cos \theta}
    \\
    &= 4\pi\frac{1}{\pi} \int_0^\infty dk \frac{1}{ik|r|}[e^{ik|r|} - e^{-ik|r|}]
    e^{-\frac{k^2}{4\alpha^2}}
    \\
    &= \frac{1}{\pi|r|} \int_{-\infty}^\infty dk \frac{1}{ik}e^{ik|r|} e^{-\frac{k^2}{4\alpha^2}}
\end{align}

Let $I(r) = \int_{-\infty}^\infty dk \frac{1}{ik}e^{ik|r|} e^{-\frac{k^2}{4\alpha}}$. Then

\begin{equation}
    \frac{dI(r)}{dr} =  \int_{-\infty}^\infty dk e^{ik|r|} e^{-\frac{k^2}{4\alpha^2}} = 2\alpha\sqrt{\pi}e^{-\alpha^2 |r|^2}
    \implies
    I(r) = 2\sqrt{\pi} \int_0^{\alpha r} e^{-t^2} dt = \pi\mathrm{erf}(\alpha r)
\end{equation}

So we have $\mathcal{F}[\tilde{V}^\ell](r) = I(r)/\pi r =  V^\ell(r)$
\footnote{We could have gotten this result by noticing the charge distribution corresponding to this potential is nothing but a Gaussian}. 
\subsection{2D Ewald}
We will now perform 2D Ewald transform. We start from the 3D Fourier transform of the long range potential.

\begin{equation}
       \tilde{V}^\ell(k) = 4\pi \frac{e^{-\frac{k^2}{4\alpha^2}}}{k^2}
\end{equation}

We can now integrate this in the $z$ direction to get 2D Fourier component:

\begin{align}
    \tilde{V}^\ell_{2D}(k) = e^{-\frac{|k|^2}{4\alpha^2}} \int_{-\infty}^\infty \frac{dk_z}{2\pi} e^{-\frac{k_z^2}{4\alpha^2}} \frac{4\pi}{k^2 + k_z^2}
\end{align}

We use the following standard integral equality:

\begin{equation}
    \int_0^\infty dk \frac{e^{-k^2}}{k^2 + t^2} = \frac{\pi \mathrm{erfc}(t)}{2t} e^{t^2}
\end{equation}

Using this, we get

\begin{equation}
     \tilde{V}^\ell_{2D}(k)
     =
     \frac{2\pi \mathrm{erfc}(\frac{k}{2\alpha})}{k}.
\end{equation}

Rather interestingly, the error function shows up in momentum space instead of real space in this case. We can also evaluate the integral of the short-range part of the interaction:

\begin{equation}
    \int d^2r \frac{\mathrm{erfc}(\alpha r)}{r} = 2\pi \int_0^\infty dr \,  \mathrm{erfc}(\alpha r) = \frac{2\sqrt{\pi}}{\alpha}
\end{equation}

The spatial limit of the interaction is easy to see. As $\mathrm{erf}(r) = 2r/\sqrt{\pi} + O(r^3)$, we have
\begin{equation}
    V^{\ell}(0) = \frac{2\alpha}{\sqrt{\pi}}.
\end{equation}

Combining these results together, the energy correction using Ewald summation is given by
\begin{equation}
\label{eq:Coulomb_ewald}
    \Delta \epsilon_C = \frac{1}{N}
    \left(
        \frac{  \rho}{2} \sum_{\v{G}\neq 0}
        \frac{2\pi\mathrm{erfc}(\frac{k}{2\alpha})}{k} -
        \frac{N}{2}\frac{2\alpha}{\sqrt{\pi}}
        -\frac{ \rho}{2} \frac{2\sqrt{\pi}}{\alpha}
        + \frac{1}{2} N \sum_{\v{R} \neq 0} \frac{\mathrm{erfc}(\alpha R)}{R}
    \right).
\end{equation}

For 2D electron gas in particular, in which the interaction is given by $\frac{2}{r_s}$, we have

\begin{equation}
    \Delta \epsilon_{2DEG} = \frac{2}{Nr_s}
    \left(
        \frac{  \rho}{2} \sum_{\v{G}\neq 0}
        \frac{2\pi\mathrm{erfc}(\frac{k}{2\alpha})}{k} -
        \frac{N}{2}\frac{2\alpha}{\sqrt{\pi}}
        -\frac{ \rho}{2} \frac{2\sqrt{\pi}}{\alpha}
        + \frac{1}{2} N \sum_{\v{R} \neq 0} \frac{\mathrm{erfc}(\alpha R)}{R}
    \right),
\end{equation}
where the $\v{r}_j$ must be chosen to have area $\pi$. These summations can then be performed easily by choosing an appropriate $\alpha$ and truncating the summation at finite $|\v G|$ and $|\v R|$, which only leaves an exponentially small error.

\section{Hartree-Fock procedure}
\label{app:SCHF}
We use standard self-consistent Hartree-Fock calculations to obtain the SCHF ground state of the electronic system. 

\subsection{Continuous translation symmetry}
\label{app:cts_HF}
In the presence of continuous symmetry, the mean-field states are fully characterized by $n(k)$, and its energies are given by 
\begin{align}
    E_{\mathrm{kin}} &= \sum_{\v k} n(\v k) \frac{|\v k|^2}{r_s^2},
\\
    E_{\mathrm{int}} &= -\frac{1}{2A} \sum_{\v k, \v k'} \tilde{V}(\v k,\v k') n(\v k) n(\v k').
\end{align}

This simple formula allows us to evaluate the energy of Fermi liquid candidate states efficiently. As our Fermi liquid ansatz, we take the following  $n(k)$:

\begin{equation}
    n(k) = 
    \begin{cases}
        0 & k < k_\mathrm{min} \\
        1 & k_\mathrm{min} \leq  k \leq k_\mathrm{max} \\
        0 & k > k_\mathrm{max}
    \end{cases}.
\end{equation}
Here $k_\mathrm{max}^2 - k_\mathrm{min}^2 = 4/a^2$ to ensure the same electron density. Setting $k_\mathrm{min} = 0$ recovers the circular Fermi surface. By minimizing the sum of kinetic and interaction energy with respect to $k_\mathrm{min}$, we can evaluate the energy of the Fermi liquid. This reduces to evaluating numerical integrals, which allows us to achieve the thermodynamic limit at a small energy cost.

\subsection{Discrete translation symmetry}

We now consider performing the self-consistent Hartree Fock while only imposing discrete translation symmetry.
As usual, we consider single-particle density matrices $P(\v{k})_{\beta\alpha}=\braket{c^\dagger_{\v k \alpha}c_{\v k \beta}}$, where $\alpha$ and $\beta$ are collective indices for valley, spin, and band. These density matrices are in one-to-one correspondence with Slater determinant states. We define the Hartree and Fock Hamiltonians as 

\begin{subequations}
\begin{align}
    h_H[P](\v k) &=\frac{1}{A}\sum_{\v g}V(\v g)\Lambda_{\v g}(\v k)\left(\sum_{\v k}\Tr[P(\v k)\Lambda_{\v g}(\v k)^\dagger]\right)\\
    h_F[P](\v k) &=-\frac{1}{A}\sum_{\v q}V(\v q)\Lambda_{\v q}(\v k) P([\v k+\v q])\Lambda_{\v q}(\v k)^\dagger
\end{align}
\end{subequations}
where $[\Lambda_{\v q}(\v k)]_{\alpha\beta}=\braket{\psi_{\v k \alpha}|e^{-i\v{q}\cdot\v{r}}|\psi_{\v{k}+\v{q} \beta}}$ are form factors, and are treated as matrices whose labels are identical to the single-particle ones. The sum over $\v{g}$ runs over reciprocal vectors while $\v{q}$ runs over all momentum transfers. Via Wick's theorem, the energy of the Slater-determinant state is
\begin{equation}
    E[P]=\frac{1}{2}\Tr[P(2h_{\textrm{kin}}+h_{H}[P]+h_F[P])]
\end{equation}
where the trace is over momentum and all other band labels. 

We use the optimal-damping algorithm (ODA) to converge to states satisfying the self-consistency condition 
\begin{equation}
    [P,h_\textrm{kin} + h_{H}[P]+h_{F}[P]]=0
\end{equation}
to tolerances approaching the square root of machine precision (i.e. machine precision in $E[P]$). We use Monkhorst-Pack grids with $12\times 12$ up to $36\times 36$ unit cells, and ensure that the range of momentum transfers $\v{q}$ considered is sufficiently large to converge the energy of the state, out to several Brillouin zone sizes. The choices of system sizes $N_k$ are always multiples of $6$ to ensure that the high-symmetry points, such as the $M$ and $K$ points when the lattice is triangular, are properly sampled. We further truncate the single-particle Hilbert spaces to $N_b = 7, 13, 19, 37$ bands, until energy convergence is reached. To avoid non-global minima, we initialize SCHF in a variety of states for each parameter point using a mix of physical ans\"atze as well as random initial states.

\section{Berry curvature and band structure of the crystalline states}
\begin{figure*}
    \centering
    \includegraphics[width=0.95\textwidth]{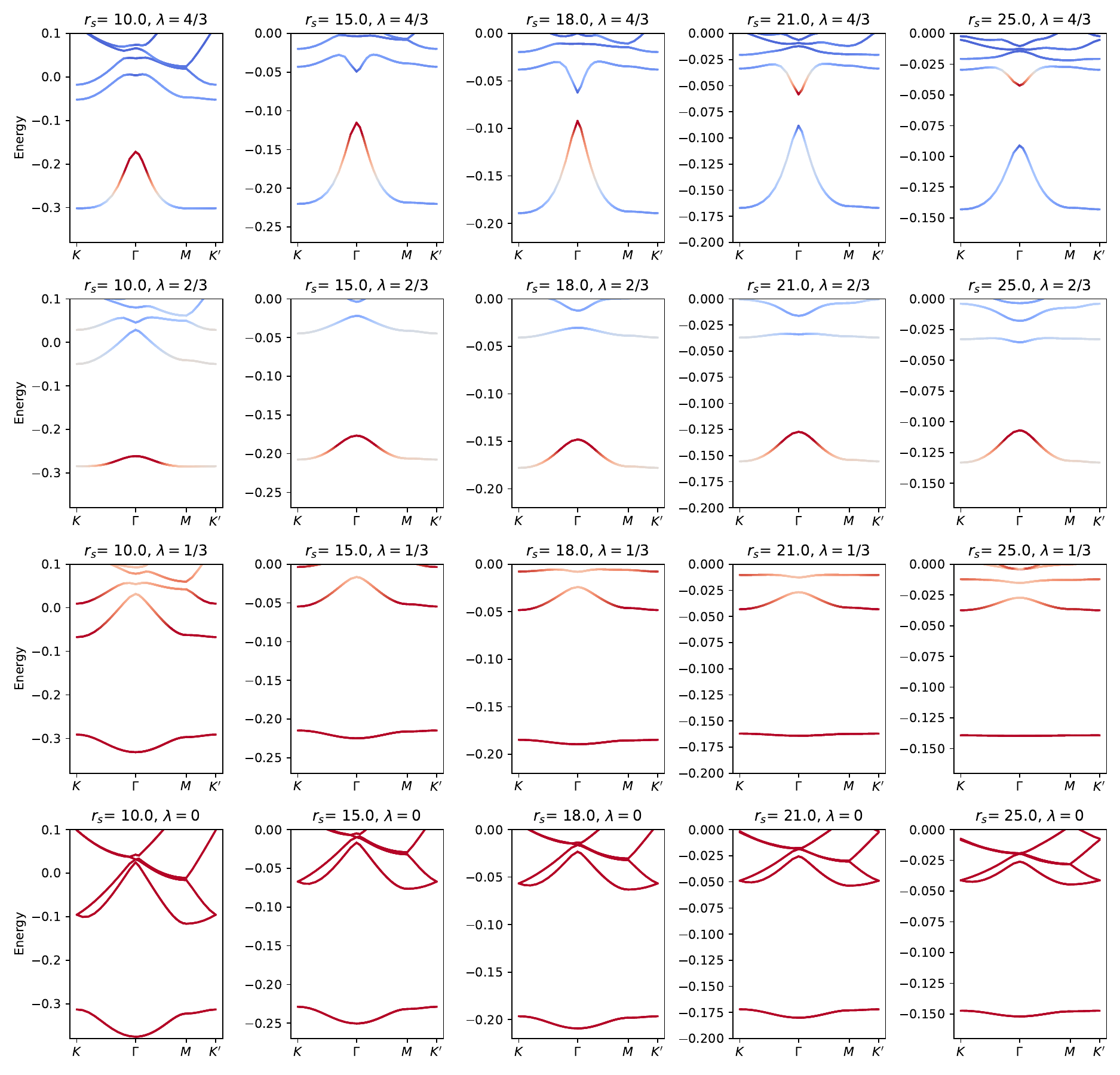}
    \caption{The band structure of the crystalline states. Colors indicate the spinor polarization $\braket{\sigma_z}$.}
    \label{fig:app_bandstructure}
\end{figure*}

\begin{figure*}
     \centering
    \includegraphics[width=0.95\textwidth]{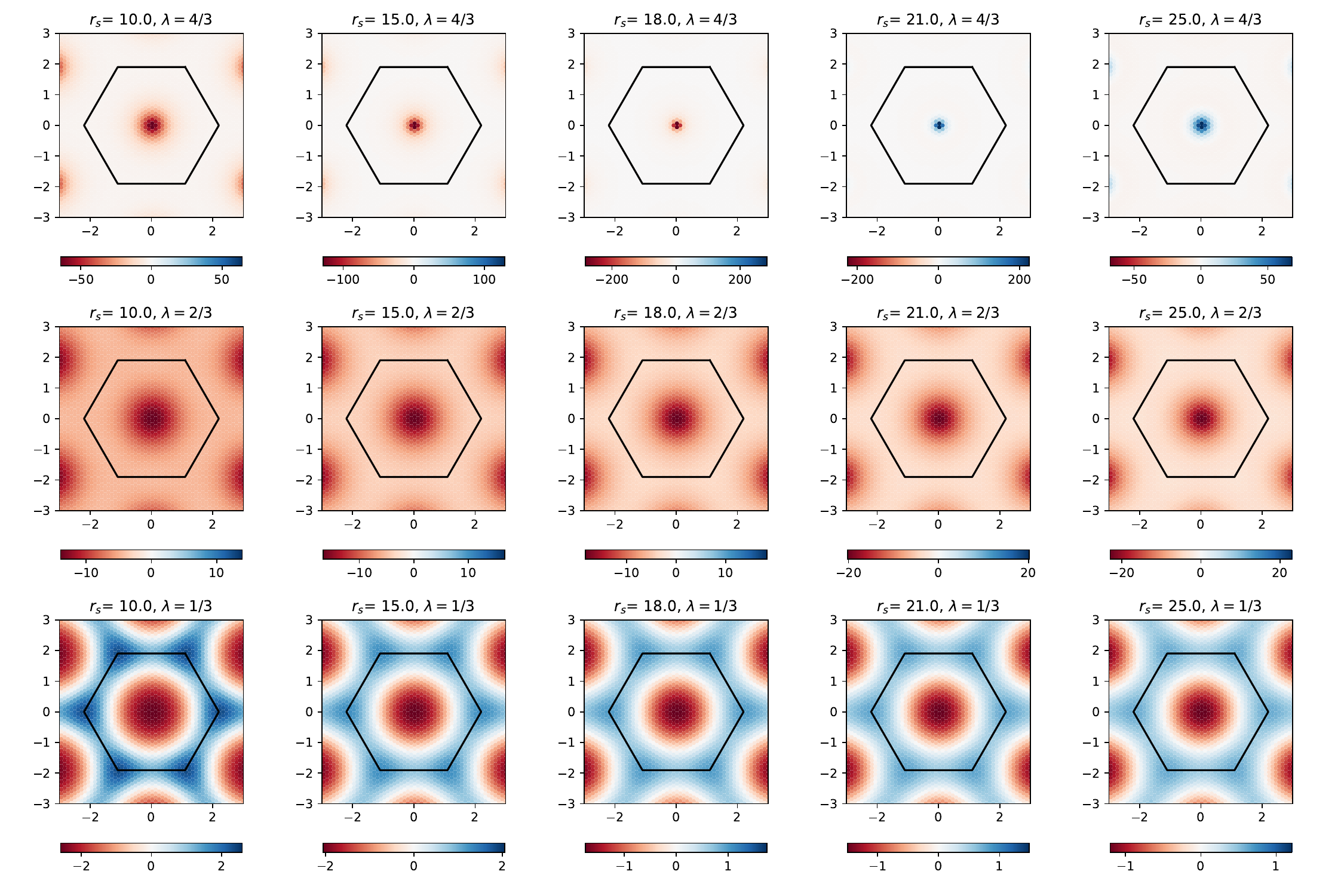}
    \caption{The Berry curvature of the crystalline states}
    \label{fig:app_berry_curvature}
\end{figure*}

In this section, we show the band structure and the Berry curvature of the crystalline states at some select values of $r_s$ and $\lambda$. In Fig.~\ref{fig:app_bandstructure}, we show the Hartree-Fock band structure of the crystalline states. The color of the plot corresponds to the polarization of the pseudospin. We can clearly see a peak at $\Gamma$ point induced by nonzero $\lambda$. In Fig.~\ref{fig:app_berry_curvature}, we show the Berry curvature of various crystalline states. We see that near the phase transition discussed in the main text, the Berry curvature becomes very non-uniform. 

\section{Finite-size scaling of the energy}
\label{app:finite-size-scaling}

In this section, we detail how we perform finite-size scaling.

As discussed in App.~\ref{app:cts_HF}, our energy evaluation is already in the thermodynamic limit, and therefore does not require finite-size scaling.

\begin{figure}
    \centering
    \subfloat[Finite-size scaling for ground state at $\lambda=2/3, r_s=11$.]{
    \includegraphics[width = 0.47\textwidth]{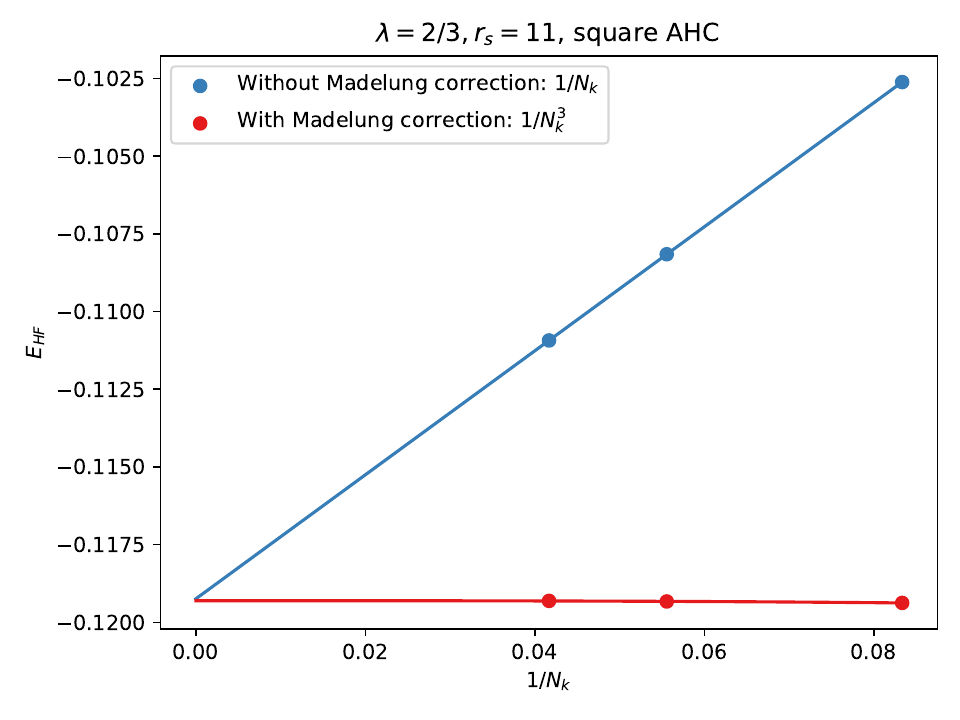}
    \label{fig:sq_finite_size}
    }
    \hfill
    \subfloat[Finite-size scaling for ground state at $\lambda=4/3, r_s=11$.]{
    \includegraphics[width = 0.47\textwidth]{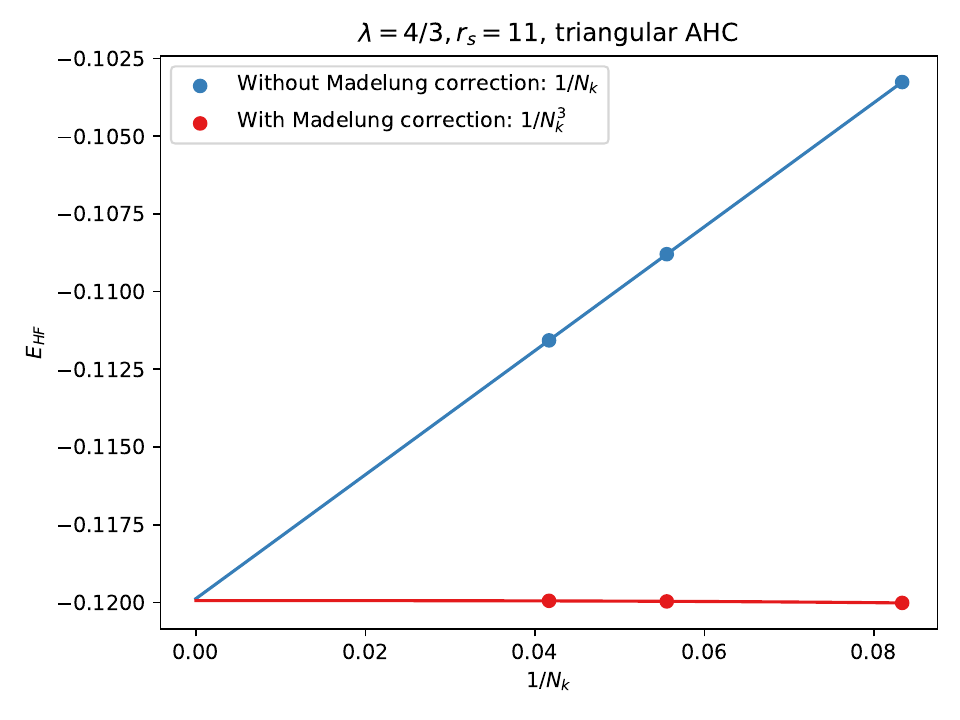}
    \label{fig:tri_finite_size}
    }
    \caption{Finite-size scaling plots for the square and triangular AHC states, with and without the Madelung correction. The Madelung correction significantly reduces the finite-size effect.}
        \label{eq:finite_size_Madelung}
\end{figure}

As for the Hartree-Fock ground states found for discrete translation symmetry, we take into account the finite-size corrections detailed in App.~\ref{app:ewald}:

\begin{equation}
    E_{\mathrm{corrected}}[P] = E[P] + \frac{2}{r_s}\Delta \epsilon_{C},
\end{equation}
where $\Delta \epsilon_C$ is given by Eq.~(\ref{eq:Coulomb_ewald}). This improves the convergence of energies greatly, from a $1/L$ asymptotic to a $1/L^3$ asymptotic~\cite{Hunt_Madelung}. In Fig.~\ref{fig:sq_finite_size},\ref{fig:tri_finite_size}, we show the result of such finite-size scaling. The inclusion of the Madelung energy correction indeed reduces the finite-size effect considerably.

\section{Semiclassical analysis for the Wigner crystal phase transition}

Here, we review the semiclassical analysis for the normal WC-halo WC transition following Ref.~\cite{joy2023wignercrystallizationbernalbilayer}. Each individual semiclassical electron can be treated as a particle with dispersion and Berry curvature specified by Eq.~(\ref{eq:ham}). On top of that, it is subjected to a harmonic potential generated by the Coulomb interactions due to all other electrons present. The electron must follow the single-particle Schrodinger's equation in momentum space~\cite{joy2023wignercrystallizationbernalbilayer,Price_Artificial_Magnetic}:

\begin{equation}
    \left( \epsilon_{\v k} + \frac{\alpha}{2}|i\v{\nabla}_{\v{k}} - \v A_{\v{k}}|^2 + \alpha \frac{\Tr g_{\mu\nu}(\v k)}{2}\right)\psi_\ell(\v k) = E_\ell \psi_\ell(\v k).
\end{equation}
Here, we approximated the confining potential by $V(\v r) = \alpha |\v r|^2/2$, $\v A_{\v{k}}$ is the Berry connection of the band, and $g_{\mu\nu}(\v k)$ is the quantum metric of the band.
When $\lambda = 0$, it reduces to the standard harmonic oscillator, whose ground state has angular momentum $\ell =0$. When $\lambda \to \infty$, the Berry curvature is $\Omega(\v{k}) = 2\pi \delta(\v{k})$, corresponding to threading $2\pi$ flux at the origin. Furthermore, $\Tr g_{\mu\nu}(\v k)$ becomes a delta function at the origin. These two effects collaborate to suppress the occupation of the $\v k = 0$ region and change the angular momentum of the ground state to $\ell = 1$. reproducing the numerical observation~\footnote{We sincerely thank Ophelia Evelyn Sommer for pointing out the quantum metric contribution to us.}.

% \bibliography{references}
% \bibliographystyle{unsrt}

% \end{document}

\end{document}